\documentclass[10pt,letterpaper]{article}
\usepackage{fix-cm}
\usepackage[textwidth=5.5in,textheight=9in,top=0.9in,centering]{geometry}
\usepackage{times}
\usepackage[round,authoryear]{natbib}

\makeatletter
\renewcommand{\normalsize}{\@setfontsize\normalsize{10pt}{11pt}}
\renewcommand{\small}{\@setfontsize\small{9pt}{10pt}}
\renewcommand{\footnotesize}{\@setfontsize\footnotesize{9pt}{10pt}}
\renewcommand{\section}{\@startsection{section}{1}{\z@}{-2ex plus -.5ex minus -.2ex}{1.5ex plus .3ex minus .2ex}{\large\bfseries\raggedright}}
\renewcommand{\subsection}{\@startsection{subsection}{2}{\z@}{-1.8ex plus -.5ex minus -.2ex}{.8ex plus .2ex}{\normalsize\bfseries\raggedright}}
\renewcommand{\subsubsection}{\@startsection{subsubsection}{3}{\z@}{-1.5ex plus -.5ex minus -.2ex}{.5ex plus .2ex}{\normalsize\bfseries\raggedright}}
\renewcommand{\paragraph}{\@startsection{paragraph}{4}{\z@}{1.5ex plus .5ex minus .2ex}{-1em}{\normalsize\bfseries}}
\renewcommand{\@maketitle}{%
  \begin{center}
    {\LARGE\bfseries\@title\par}
    \vspace{8pt}
    {\begin{tabular}[t]{c}\@author\end{tabular}}
  \end{center}
  \vspace{4pt}%
}
\makeatother
\normalsize

\usepackage[T1]{fontenc}
\usepackage[utf8]{inputenc}
\usepackage{amsmath,amssymb,booktabs,array,multirow,graphicx,float}
\usepackage{microtype}
\usepackage{xcolor}
\usepackage[hyphens]{url}
\usepackage{hyperref}
\hypersetup{colorlinks=true,citecolor=blue,linkcolor=blue,urlcolor=blue,
  pdftitle={RandSlot: Learning Compact Visual Document Representations with Random Soft Tokens},
  pdfauthor={Dewen Guo, Shi Yu, Lingxiao Zhang, Yang Zhang, Tao XU, Dan Wang}}
\newcommand{\method}{RandSlot}
\newcommand{\normop}{\operatorname{N}}

\title{\LARGE\bfseries RandSlot: Learning Compact Visual Document
Representations with Random Soft Tokens}
\author{%
  \normalsize Dewen Guo\textsuperscript{1}, Shi Yu\textsuperscript{2},
  Lingxiao Zhang\textsuperscript{1}, Yang Zhang\textsuperscript{1},
  Tao XU\textsuperscript{1}, Dan Wang\textsuperscript{1}\\[6pt]
  \normalsize\textsuperscript{1}Ant Group\qquad\textsuperscript{2}Tsinghua University\\[4pt]
  \normalsize\href{mailto:gdw4395@gmail.com}{\texttt{gdw4395@gmail.com}}%
}
\date{}
\begin{document}
\maketitle
\begin{abstract}
Visual document retrieval requires expressive representations to match queries with evidence distributed across text, tables, and page layouts.
Multi-vector representations capture fine-grained information, but storing and comparing many vectors introduces substantial retrieval costs.
In this paper, we introduce \method, a simple approach to learn compact visual document representations with random soft tokens.
During training, we append independently-sampled random unit vectors to query and document input sequences and resample them at every use, without introducing learnable soft-token parameters.
The encoder contextualizes these auxiliary inputs with the original content to produce a small set of retrieval vectors.
A standard late-interaction objective trains the encoder to extract relevant information under varying input conditions.
Experiments with different backbone models show that \method\ improves retrieval quality over alternative readout strategies under the same vector budget.
Further analysis shows that these gains can persist when random soft tokens are replaced with zeros at inference, demonstrating that random inputs during training can improve compact retrieval representations even when inference no longer requires sampling.
\end{abstract}

\section{Introduction}
\label{sec:introduction}

Visual document retrieval matches text queries to evidence distributed across text, tables, charts, and page layouts.
A financial query, for example, may require interpreting a value together with its row heading and accompanying footnote.
Encoding document pages as images preserves these relationships, while multi-vector representations support fine-grained matching between queries and page content \citep{ma2024dse,faysse2024colpali}.
However, storing and comparing large sets of vectors increases retrieval costs.
Under a compact representation budget, the encoder must organize the relevant information into only a few outputs.
The challenge is therefore not simply to reduce the number of vectors, but to learn a small set of representations that jointly support effective retrieval.

Existing approaches address this challenge through post-encoding compression or direct construction of compact representations.
Light-ColPali and DocPruner merge or prune page embeddings \citep{ma2025storage,yan2025docpruner}, whereas MetaEmbed introduces learnable tokens and CausalEmbed autoregressively feeds model-generated representations back into the input \citep{xiao2026metaembed,huo2026causalembed}.
These approaches highlight a design choice beyond the number of output vectors: how to construct the auxiliary inputs used to produce them.
Learnable tokens provide persistent parameters that adapt during training, while autoregressive feedback supplies content-dependent inputs.
This raises a question: \emph{must auxiliary inputs be learned or carry content to produce effective compact retrieval representations?}

An auxiliary input and its resulting output serve different roles.
An appended token can attend to the preceding content, so its contextualized output can encode information that is absent from the input token itself.
Even a token without predefined semantics can therefore provide a position at which the encoder gathers information for retrieval.
This motivates examining whether the values assigned to these inputs need to remain fixed or be optimized.
Inspired by stochastic representation learning and random embedding injection \citep{gao2021simcse,jain2023neftune,kim2026rsp}, we investigate continuously resampling the auxiliary inputs during training.
Our intuition is that varying these inputs while retaining the same retrieval objective encourages the encoder to extract useful content under changing input conditions, potentially reducing its dependence on particular soft-token values.

We introduce \method, a simple approach to learning compact visual document representations with random soft tokens.
For each query and document, \method\ sequentially appends independently sampled random vectors, normalized to unit length to control their scale.
The encoder produces an initial vector from the original input and an additional output vector for each appended token.
These content-dependent outputs form the retrieval representation and are jointly optimized with a standard late-interaction contrastive objective.
Random soft tokens are resampled for every encoding call and introduce no additional trainable token parameters.
Although the auxiliary inputs are random, the output representations are trained to capture information relevant to query--document matching.

Experiments on ViDoRe V1--V3 demonstrate the effectiveness of this approach: \method-3B exceeds the published CausalEmbed results while using substantially fewer retrieval vectors.
To isolate the role of input construction, we compare random, zero, fixed-random, learned, and direct-feedback inputs under the same training procedure and vector budget.
\method\ achieves the highest average performance on ViDoRe V3 with both evaluated backbones.
Replacing its random inputs with zeros at inference preserves most of the performance, supporting the value of stochastic inputs during training.
Further analyses examine the complementary contributions of document output vectors and show how retrieval quality changes with the inference vector budget and its associated costs.

Our contributions are threefold:
\begin{itemize}
    \item We propose \method, which learns compact visual document representations with resampled random soft tokens and no additional trainable token parameters.
    \item Among five auxiliary-input strategies under matched training settings and vector budgets, \method\ achieves the highest average ViDoRe V3 performance with both backbones.
    \item We show that gains persist with zero-input inference, analyze document-readout complementarity via MaxSim assignments and readout ablations, and evaluate quality--cost trade-offs across inference budgets.
\end{itemize}
\section{Related work}
\label{sec:related}
\paragraph{Multimodal and visual document retrieval.}
Multimodal embedding models map heterogeneous inputs into a shared representation space. CLIP establishes cross-modal alignment through image--text contrastive learning \citep{radford2021clip}, while DSE encodes document screenshots to incorporate visual information into retrieval \citep{ma2024dse}. More recent methods, including MM-Embed, GME, VLM2Vec-V2, and Qwen3-VL-Embedding, adapt vision--language models for retrieval across text, images, and visual documents \citep{lin2025mmembed,zhang2025gme,meng2025vlm2vecv2,li2026qwen3vlembedding}. These developments provide a foundation for learning expressive document representations, with increasing attention to their storage and retrieval costs.

\paragraph{Compact multi-vector retrieval.}
ColBERT and ColPali preserve fine-grained matching through late interaction between token-level representations \citep{khattab2020colbert,faysse2024colpali}. To reduce the resulting storage costs, Light-ColPali merges patch embeddings, DocPruner adaptively prunes them, and ReinPool learns to select and pool embeddings \citep{ma2025storage,yan2025docpruner,cha2026reinpool}. ColBERTv2 compresses vector representations, while XTR and MUVERA improve search efficiency through token retrieval and fixed-dimensional encodings, respectively \citep{santhanam2022colbertv2,lee2023xtr,dhulipala2024muvera}. \method\ addresses compactness during representation construction by training the encoder to produce a small set of retrieval vectors.

\paragraph{Soft tokens and continuous-state readout.}
Learned latent queries aggregate information in Perceiver and BLIP-2, while prefix tuning and prompt tuning adapt pretrained models through continuous inputs \citep{jaegle2021perceiver,li2023blip2,li2021prefixtuning,lester2021prompttuning}. Pause Tokens further show that additional token positions can support computation before prediction \citep{goyal2024pausetokens}. For retrieval, MetaEmbed appends learnable Meta Tokens and uses their contextualized hidden states as compact embeddings. Drawing on Matryoshka Representation Learning, it jointly optimizes nested vector groups to support multiple retrieval budgets \citep{kusupati2022mrl,xiao2026metaembed}. SLQ learns shared latent queries while keeping the multimodal backbone frozen \citep{lou2026slq}. Another line reuses intermediate hidden states: Coconut supports continuous reasoning, GIRCSE iteratively refines text embeddings, and CausalEmbed autoregressively generates visual document embeddings \citep{hao2025coconut,tsai2025gircse,huo2026causalembed}. \method\ constructs its auxiliary inputs from independently resampled random unit vectors and trains the encoder adaptation parameters to extract retrieval representations at these positions.

\paragraph{Stochastic inputs and representation learning.}
Noise injection and dropout provide established approaches to regularizing neural networks \citep{bishop1995noise,srivastava2014dropout}. SimCSE uses dropout to construct contrastive views, while R-Drop encourages consistency between predictions under different dropout masks \citep{gao2021simcse,liang2021rdrop}. At the input level, NEFTune perturbs token embeddings during instruction fine-tuning \citep{jain2023neftune}. Random Soft Prompts append sampled embeddings to diversify language-model reasoning; their application to reinforcement learning also uses random prompts during training and removes them at evaluation \citep{kim2026rsp}. \method\ studies stochastic inputs for compact retrieval representation learning. Random soft tokens occupy auxiliary readout positions, and the encoder is trained with a late-interaction contrastive objective over their contextualized outputs and the initial content representation.

\section{Method}
\label{sec:method}
\begin{figure}[!t]
  \centering
  \includegraphics[width=\linewidth]{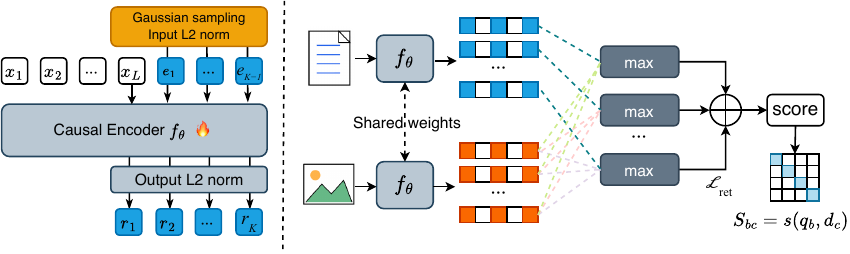}
  \caption{Overview of \method. Left: $K-1$ random unit vectors are appended to the content. Position $x_L$ supplies $\mathbf r_1$, and the appended positions supply $\mathbf r_2,\ldots,\mathbf r_K$; all outputs are L2-normalized. Right: queries and pages share encoder weights but sample inputs independently. MaxSim yields the score matrix for contrastive training, with positive pairs highlighted.}
  \label{fig:overview}
\end{figure}

\method\ encodes queries and document pages into compact representations by appending random soft tokens to the original input. Figure~\ref{fig:overview} shows how a shared encoder produces contextualized readouts and a late-interaction objective trains them for retrieval.

\subsection{Random soft-token readout}
Let $f_\theta$ be a shared causal encoder that produces $K_q$ vectors for a query $q$ and $K_d$ vectors for a document page $d$. For either input $x$, let $X_x$ denote the embedding sequence containing its prompt and textual or visual content. To obtain $K_x$ readouts, we append $K_x-1$ random soft tokens $E_x=(\mathbf e_1,\ldots,\mathbf e_{K_x-1})$, where
\begin{equation}
\mathbf z_t\sim\mathcal N(\mathbf0,I_{D_{\rm in}}),\qquad
\mathbf e_t=\normop(\mathbf z_t),\qquad t=1,\ldots,K_x-1.
\label{eq:random}
\end{equation}
Here $D_{\rm in}$ is the encoder input dimension and $\normop(\mathbf v)=\mathbf v/\max(\|\mathbf v\|_2,\varepsilon)$. Unit normalization fixes the input scale. The draws are independent across examples, query and page inputs, and appended positions. Within an encoding, each new position receives one draw and all previously appended vectors remain unchanged; a new encoding draws a fresh set.

The encoder reads the last-layer hidden state at the final valid position of each extended prefix:
\begin{align}
\mathbf h_t(x;E_x)&=\operatorname{Last}\!\left(f_\theta([X_x;\mathbf e_1;\ldots;\mathbf e_{t-1}])\right),\label{eq:readout}\\
\mathbf r_t(x;E_x)&=\normop(\mathbf h_t(x;E_x)),\qquad t=1,\ldots,K_x.\label{eq:output}
\end{align}
For $t=1$, the prefix contains only $X_x$, giving the original-content readout. Each appended position adds a readout with causal access to the content and preceding soft tokens. The representation $R_\theta(x;E_x)=[\mathbf r_1,\ldots,\mathbf r_{K_x}]$ therefore contains one original-content and three appended readouts at the default four-vector budget. Output normalization makes retrieval dot products cosine similarities. 

\subsection{Retrieval training}
We independently encode a query and page as $Q=R_\theta(q;E_q)$ and $D=R_\theta(d;E_d)$. Their relevance is measured by directional MaxSim:
\begin{equation}
s_\theta(q,d)=\sum_{i=1}^{K_q}\max_{1\le j\le K_d}\mathbf q_i^\top\mathbf d_j,
\label{eq:maxsim}
\end{equation}
where $\mathbf q_i$ and $\mathbf d_j$ are the normalized readouts in $Q$ and $D$. Each query vector selects its strongest page-vector match, allowing evidence to align across different readout positions.

For a training batch $\mathcal B$ of $B$ queries, we gather document representations across devices to form $M$ candidates. The scores $S_{bc}=s_\theta(q_b,d_c)$ form the matrix shown in Figure~\ref{fig:overview}. With $\pi(b)$ denoting the positive document index for query $b$, we apply a query-to-document InfoNCE loss \citep{oord2018cpc}:
\begin{equation}
\mathcal L_{\rm ret}=-\frac1B\sum_{b=1}^{B}\log
\frac{\exp(S_{b,\pi(b)}/\tau)}
{\sum_{c=1}^{M}\exp(S_{bc}/\tau)},
\label{eq:loss}
\end{equation}
where $\tau$ is the temperature. All readouts are trained jointly through the query--page scores.

The soft tokens have no trainable parameters. Training updates the encoder's adaptation parameters $\theta$, optimizing the retrieval objective over batches from the training set $\mathcal T$ and independently sampled query and page inputs:
\begin{equation}
\min_\theta\;\mathbb E_{\mathcal B\sim\mathcal T}\,
\mathbb E_{E_q,E_d}\left[\mathcal L_{\rm ret}(\theta;\mathcal B,E_q,E_d)\right].
\label{eq:expectation}
\end{equation}
Each encoding uses one sampled soft-token sequence. 

\subsection{Inference}
Pages are encoded independently in advance and stored as $K_d$ vectors each. At query time, the encoder produces $K_q$ vectors and ranks candidate pages by MaxSim (Eq.~\eqref{eq:maxsim}).

The same trained encoder supports random inputs (\textbf{RS-R}) or zero inputs (\textbf{RS-Z}) at the appended positions. RS-R follows the training sampling rule; RS-Z preserves the positions, attention structure, and output budget without retraining. Varying the number of appended positions adjusts the inference budget, as evaluated in Section~\ref{sec:budget}. Appendix~\ref{app:implementation} details sequential execution and caching.

\section{Experiments}
\label{sec:experiments}
\subsection{Experimental setup}
\paragraph{Data and evaluation.}
We train on the full training split of the ColPali dataset and evaluate on ten ViDoRe V1 tasks, four V2 tasks, and eight public V3 tasks \citep{faysse2024colpali,mace2025vidorev2,loison2026vidorev3}. The two private V3 datasets are excluded. We report page-level nDCG@5 as percentages, with per-task scores and version averages; absolute differences are in percentage points (pp). The main comparison covers all three versions; ablations focus on V3's professional documents and human-verified queries involving visual evidence, comparison, and multi-hop retrieval.

\paragraph{Models and training.}
We instantiate \method-3B and \method-4B with Qwen2.5-VL-3B and Qwen3-VL-4B (3B/4B in tables) \citep{bai2025qwen25vl,bai2025qwen3vl}. Both output four vectors per side, with dimensions of 2,048 and 2,560, respectively. We train both for one epoch (923 steps) on eight NVIDIA A100 80GB GPUs, using a global batch of 128 (16 per device). Shared settings include LoRA rank 16, scaling 32, learning rate $2\times10^{-4}$, and a 1,536 visual-token limit \citep{hu2021lora}. Appendix~\ref{app:implementation} gives the initialization, templates, and remaining settings.

\paragraph{Comparison protocol.}
We use the published results and budgets for Base (full-token), Random-Select, SemCluster, K-Means, 1D-Pooling, BiQwen2, and CausalEmbed from \citet{huo2026causalembed}, preserving their training recipes and evaluation protocol. Random-Select samples patch vectors; CausalEmbed is CausalQwen with 16 query and 32 page vectors. Our controls share the initialization, readout, retrieval objective, training recipe, and four-vector budget, changing only appended inputs. Learned-Feedback (L) and Direct-Feedback (DF) implement MetaEmbed-style learnable tokens and CausalEmbed-style autoregressive feedback, respectively. We retain their respective input construction mechanisms while aligning the training procedure, hyperparameters, and retrieval-vector budget with RandSlot to ensure a fair comparison.

\subsection{Overall retrieval performance}
\begin{table}[!t]
\centering
\caption{Retrieval performance on ViDoRe V3 and V2 (nDCG@5 $\uparrow$, \%), with task scores and averages reported separately for each version.}
\label{tab:main-v32}
\begingroup
\fontsize{7.7}{9.2}\selectfont
\setlength{\tabcolsep}{1.3pt}
\begin{tabular*}{\linewidth}{@{\extracolsep{\fill}}lrrrrrrrrrrrrrrr@{}}
\toprule
& & \multicolumn{9}{c}{ViDoRe V3} & \multicolumn{5}{c}{ViDoRe V2} \\ \cmidrule(lr){3-11}\cmidrule(lr){12-16}
Method & $K_d$ & HR & Fin-E & Ind. & Phar. & C.S. & Ener. & Phys. & Fin-F & Avg & ESG & Bio & Econ & ESG-H & Avg \\
\midrule
Base$^\dagger$ & 4962 & 47.3 & 50.0 & 41.6 & 56.1 & 68.6 & 57.1 & 43.4 & 37.5 & 50.2 & 54.9 & 59.1 & 54.4 & 62.0 & 57.6 \\
Random-Select$^\dagger$ & 32 & 24.8 & 25.0 & 20.3 & 38.4 & 49.2 & 34.8 & 33.8 & 14.4 & 30.1 & 29.3 & 42.5 & 35.4 & 30.4 & 34.4 \\
SemCluster$^\dagger$ & 32 & 35.9 & 36.6 & 30.7 & 48.1 & 59.4 & 45.3 & 39.0 & 26.7 & 40.2 & 42.6 & 51.5 & 48.5 & 39.5 & 45.5 \\
K-Means$^\dagger$ & 32 & 36.5 & 36.0 & 30.9 & 46.7 & 57.9 & 46.4 & 38.3 & 24.7 & 39.7 & 41.3 & 50.4 & 53.0 & 35.9 & 45.2 \\
1D-Pooling$^\dagger$ & 32 & 26.0 & 28.3 & 23.7 & 40.4 & 56.1 & 39.7 & 37.9 & 18.6 & 33.8 & 25.1 & 47.2 & 41.7 & 26.1 & 35.0 \\
BiQwen2$^\dagger$ & 1 & 33.6 & 33.3 & 23.6 & 46.0 & 55.0 & 39.9 & 38.3 & 20.0 & 36.2 & 41.3 & 48.7 & 49.3 & 50.4 & 46.4 \\
CausalEmbed$^\dagger$ & 32 & 42.9 & 39.5 & 32.7 & 48.3 & 64.3 & 46.0 & 39.0 & 28.5 & 42.6 & 49.8 & 50.5 & 52.2 & 53.7 & 51.6 \\
\midrule
\textbf{RS (3B)} & 4 & 44.1 & 44.0 & 35.3 & 49.1 & 67.3 & 49.8 & 43.6 & 29.8 & 45.4 & 48.9 & 53.9 & 55.6 & 53.6 & 53.0 \\
\textbf{RS (4B)} & 4 & 44.6 & 44.9 & 35.2 & 51.2 & 70.8 & 49.9 & 44.8 & 27.1 & 46.1 & 51.0 & 56.0 & 58.4 & 48.8 & 53.5 \\
\bottomrule
\end{tabular*}
\endgroup
\par\smallskip
\begin{minipage}{\linewidth}\footnotesize RS: RandSlot. $\dagger$: published results from \citet{huo2026causalembed}. Page-vector counts retain the reported budgets; Base counts are token statistics. Query/page budgets are 4/4 for RandSlot, 16/32 for CausalEmbed, and 1/1 for BiQwen2. Other query counts are not specified in the source table.\end{minipage}
\end{table}

\begin{table}[!t]
\centering
\caption{Retrieval performance on the ten ViDoRe V1 tasks (nDCG@5 $\uparrow$, \%).}
\label{tab:main-v1}
\begingroup
\footnotesize
\setlength{\tabcolsep}{2pt}
\begin{tabular}{@{}lrrrrrrrrrrrr@{}}
\toprule
Method & $K_d$ & Arxiv & Doc & Info & TabF & TatD & Shift & Syn-AI & Syn-En & Syn-GR & Syn-HI & Avg \\
\midrule
Base$^\dagger$ & 5046 & 87.6 & 62.2 & 93.1 & 87.4 & 80.5 & 84.7 & 98.3 & 96.0 & 95.5 & 99.3 & 88.5 \\
Random-Select$^\dagger$ & 32 & 72.1 & 41.4 & 74.4 & 84.9 & 68.5 & 48.8 & 85.4 & 86.7 & 83.8 & 84.4 & 73.0 \\
SemCluster$^\dagger$ & 32 & 83.2 & 52.9 & 85.1 & 87.7 & 70.5 & 65.4 & 93.3 & 91.4 & 90.2 & 93.8 & 81.3 \\
K-Means$^\dagger$ & 32 & 83.7 & 52.4 & 86.1 & 86.6 & 67.2 & 72.5 & 93.2 & 92.1 & 88.9 & 92.8 & 81.6 \\
1D-Pooling$^\dagger$ & 32 & 73.0 & 43.2 & 82.9 & 79.1 & 61.4 & 62.5 & 88.3 & 87.3 & 87.3 & 93.1 & 75.8 \\
BiQwen2$^\dagger$ & 1 & 83.3 & 51.6 & 82.9 & 83.3 & 66.1 & 72.3 & 94.0 & 86.1 & 92.5 & 95.2 & 80.7 \\
CausalEmbed$^\dagger$ & 32 & 80.7 & 55.3 & 85.7 & 82.5 & 62.0 & 73.7 & 92.7 & 90.7 & 93.5 & 94.1 & 81.1 \\
\midrule
\textbf{RS (3B)} & 4 & 85.4 & 57.3 & 88.3 & 91.0 & 69.5 & 76.1 & 96.5 & 93.5 & 94.0 & 95.0 & 84.7 \\
\textbf{RS (4B)} & 4 & 86.7 & 61.5 & 89.7 & 92.4 & 71.5 & 78.4 & 96.5 & 94.1 & 92.3 & 96.3 & 85.9 \\
\bottomrule
\end{tabular}
\endgroup
\par\smallskip
\begin{minipage}{\linewidth}\footnotesize RS: RandSlot. Sources and budget conventions follow Table~\ref{tab:main-v32}.\end{minipage}
\end{table}

With Qwen2.5-VL-3B, \method\ exceeds the published CausalEmbed results on all three ViDoRe versions while using substantially fewer retrieval vectors (Tables~\ref{tab:main-v32} and~\ref{tab:main-v1}). On V3, its average nDCG@5 reaches 45.4, compared with 42.6 for CausalEmbed; the corresponding improvements on V2 and V1 are 1.4 and 3.6 points. These gains are achieved with 4 query and 4 page vectors, compared with CausalEmbed's 16 and 32, respectively---a reduction of 87.5\% in page-vector count. The advantage extends across 20 of the 22 tasks, including all eight V3 tasks. 

Using Qwen3-VL-4B further improves the averages across all three versions, reaching 46.1 on V3. The full-token baseline retains higher average scores, leaving room to further improve compact representations. We next examine the contribution of random soft tokens through controlled comparisons at matched vector budgets.

\subsection{Effect of the feedback strategy}
\label{sec:ablations}
\begin{figure}[!t]
  \centering
  \includegraphics[width=\linewidth]{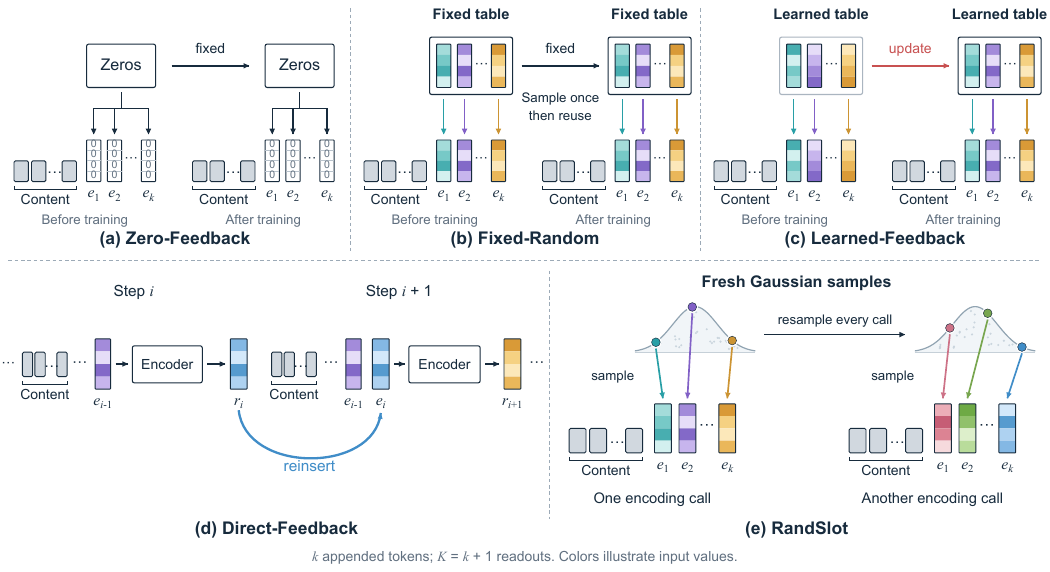}
  \caption{Input construction for five strategies. (a--c) Training leaves zero/fixed inputs unchanged and updates learned tables. (d) Direct-Feedback reinserts $\mathbf r_i$ as $\mathbf e_i$. (e) RandSlot independently resamples inputs on every encoding call, including repeated calls on the same query or page. All variants train the encoder; F/L use separate query/page tables. Here $k=K_x-1$. Normalization is omitted; see Eq.~\ref{eq:random} and Appendix~\ref{app:implementation}.}
  \label{fig:feedback-strategies}
\end{figure}


Figure~\ref{fig:feedback-strategies} illustrates five input strategies with the same output structure. \textbf{Zero-Feedback (Z)} appends zero vectors. \textbf{Fixed-Random (F)} appends random unit vectors sampled once and kept fixed throughout training and inference. \textbf{Learned-Feedback (L)} uses trainable input vectors and applies L2 normalization before use. F and L use separate sets of position-specific vectors for queries and documents. \textbf{Direct-Feedback (DF)} feeds the latest normalized output vector back as the next input, $\mathbf e_t=\mathbf r_t$, while preserving gradient flow. \textbf{RandSlot (R)} independently samples new random unit vectors for each encoding call (Eq.~\ref{eq:random}). All strategies train the encoder. 

\begin{table}[!t]
\centering
\caption{Feedback strategies on the eight public ViDoRe V3 datasets (nDCG@5 $\uparrow$, \%), with four vectors per query and page. Bold marks the best result per backbone and column. Learned-Feedback uses separate query and page tables.}
\label{tab:feedback}
\begingroup
\footnotesize
\setlength{\tabcolsep}{2pt}
\begin{tabular*}{\linewidth}{@{\extracolsep{\fill}}llrrrrrrrrr@{}}
\toprule
Backbone & Feedback & HR & Fin-E & Ind. & Phar. & C.S. & Ener. & Phys. & Fin-F & Avg \\
\midrule
3B & Zero-Feedback & 41.7 & 42.0 & 32.7 & 49.4 & 65.1 & 48.5 & 41.8 & \textbf{30.4} & 43.9 \\
3B & Fixed-Random & 42.5 & 41.9 & 33.7 & 47.4 & 64.5 & 48.0 & \textbf{43.6} & 28.4 & 43.8 \\
3B & Learned-Feedback & 42.2 & 37.3 & 31.9 & \textbf{50.3} & 64.2 & 48.3 & 42.0 & 27.0 & 42.9 \\
3B & Direct-Feedback & 40.8 & 41.4 & 30.0 & 48.4 & 62.1 & 45.7 & 40.0 & 28.2 & 42.1 \\
3B & RandSlot & \textbf{44.1} & \textbf{44.0} & \textbf{35.3} & 49.1 & \textbf{67.3} & \textbf{49.8} & \textbf{43.6} & 29.8 & \textbf{45.4} \\
\midrule
4B & Zero-Feedback & 43.9 & \textbf{45.8} & 32.0 & \textbf{52.9} & 66.2 & 48.8 & 43.6 & 27.9 & 45.1 \\
4B & Fixed-Random & 43.1 & 40.2 & 34.7 & 51.5 & 70.1 & \textbf{49.9} & 43.5 & 27.8 & 45.1 \\
4B & Learned-Feedback & 42.6 & 40.0 & 33.2 & 51.4 & 69.2 & 47.9 & 43.7 & 27.4 & 44.4 \\
4B & Direct-Feedback & 44.0 & 45.1 & 33.7 & 51.9 & 67.5 & 49.5 & 43.1 & \textbf{28.7} & 45.4 \\
4B & RandSlot & \textbf{44.6} & 44.9 & \textbf{35.2} & 51.2 & \textbf{70.8} & \textbf{49.9} & \textbf{44.8} & 27.1 & \textbf{46.1} \\
\bottomrule
\end{tabular*}
\endgroup
\end{table}

\paragraph{Beyond additional readout positions.}
Zero-Feedback retains all readout positions, testing whether the gain comes simply from having more outputs. \method\ improves the V3 average from 43.9 to 45.4 with Qwen2.5-VL-3B and from 45.1 to 46.1 with Qwen3-VL-4B (Table~\ref{tab:feedback}). For the 3B comparison, three runs with seeds 42, 123, and 2024 yield $43.50\!\pm\!0.43$ for Zero-Feedback and $45.03\!\pm\!0.40$ for \method, preserving a $1.53$-point average gain (Appendix~\ref{app:seed-robustness}). These gains show that the choice of appended inputs matters even when the number of retrieval vectors is unchanged.

\paragraph{Resampling versus a fixed initialization.}
Fixed-Random tests whether a single random initialization is sufficient to obtain the benefit. Resampling improves the V3 average over fixed inputs by 1.6 points with Qwen2.5-VL-3B and 1.0 with Qwen3-VL-4B. Fixed-Random itself remains close to Zero-Feedback with the former backbone and matches it with the latter. A fixed nonzero input therefore does not account for the advantage of resampling.

\paragraph{Random versus learnable-token feedback.}
Learned-Feedback tests whether optimizing persistent input vectors improves the readout. \method\ exceeds it by 2.5 points with Qwen2.5-VL-3B and 1.7 with Qwen3-VL-4B. Learned-Feedback also scores below Fixed-Random with both backbones despite using the same table structure. The highest V3 average with each backbone thus comes from resampled inputs without additional learned feedback parameters.

\paragraph{Random versus autoregressive feedback.}
Direct-Feedback reaches V3 averages of 42.1 with Qwen2.5-VL-3B and 45.4 with Qwen3-VL-4B. \method\ improves on these scores by 3.3 and 0.7 points, winning 8 and 5 tasks, respectively. Although Direct-Feedback outperforms zero, fixed-random, and learned inputs with Qwen3-VL-4B, random soft tokens achieve the highest average with both backbones.
The gains of Direct-Feedback over Zero-Feedback vary across backbones, suggesting that feeding model-derived representations back into the input does not consistently improve retrieval. This echoes LaSER's observation that contrastively trained latent reasoning yields uneven gains across architectures \citep{jin2026laser}.

\paragraph{Variation across tasks and backbones.}
Relative to Zero-Feedback, C.S. and Industrial improve with both backbones, whereas Pharmaceuticals and Fin-F decline. Fin-E improves by 2.0 points with Qwen2.5-VL-3B but declines by 0.9 with Qwen3-VL-4B. The larger backbone achieves higher absolute quality but a smaller gain over zero feedback. Appendix~\ref{app:tasks} provides V3 task scores and differences.

\subsection{Separating training from inference feedback}
\label{sec:train-infer}
\begin{table}[!t]
\centering
\caption{Training and inference feedback on ViDoRe V3 (nDCG@5 $\uparrow$, \%), with four vectors per side. RS-R and RS-Z use the same randomly trained encoder; Z uses zero feedback throughout.}
\label{tab:train-infer}
\begingroup
\footnotesize
\setlength{\tabcolsep}{1.4pt}
\begin{tabular*}{\linewidth}{@{\extracolsep{\fill}}lllrrrrrrrrr@{}}
\toprule
Backbone & Training & Inference & HR & Fin-E & Ind. & Phar. & C.S. & Ener. & Phys. & Fin-F & Avg \\
\midrule
3B & Random & Random (RS-R) & 44.1 & 44.0 & 35.3 & 49.1 & 67.3 & 49.8 & 43.6 & 29.8 & 45.4 \\
3B & Random & Zero (RS-Z) & 44.1 & 44.2 & 35.0 & 49.0 & 67.5 & 49.8 & 43.6 & 29.7 & 45.4 \\
3B & Zero & Zero (Z) & 41.7 & 42.0 & 32.7 & 49.4 & 65.1 & 48.5 & 41.8 & 30.4 & 43.9 \\
\midrule
4B & Random & Random (RS-R) & 44.6 & 44.9 & 35.2 & 51.2 & 70.8 & 49.9 & 44.8 & 27.1 & 46.1 \\
4B & Random & Zero (RS-Z) & 44.8 & 44.4 & 34.7 & 50.9 & 70.5 & 49.2 & 44.6 & 26.9 & 45.8 \\
4B & Zero & Zero (Z) & 43.9 & 45.8 & 32.0 & 52.9 & 66.2 & 48.8 & 43.6 & 27.9 & 45.1 \\
\bottomrule
\end{tabular*}
\endgroup
\end{table}

We replace random soft tokens with zeros in the same trained encoder, retaining all 4 outputs. The key comparison uses zero inputs at inference for both randomly trained and zero-trained models (Table~\ref{tab:train-infer}). With Qwen2.5-VL-3B, \method\ retains its V3 average of 45.4, exceeding the zero-trained model's 43.9. With Qwen3-VL-4B, zero inputs lower the average from 46.1 to 45.8, still above the zero-trained model's 45.1. These retained gains under matched inference inputs show that stochastic training improves the learned encoder without requiring continued sampling at inference.

\section{Analysis}
\label{sec:analysis}
\subsection{Effect of the inference vector budget}
\label{sec:budget}
\begin{figure}[!t]
\centering
\includegraphics[width=\linewidth]{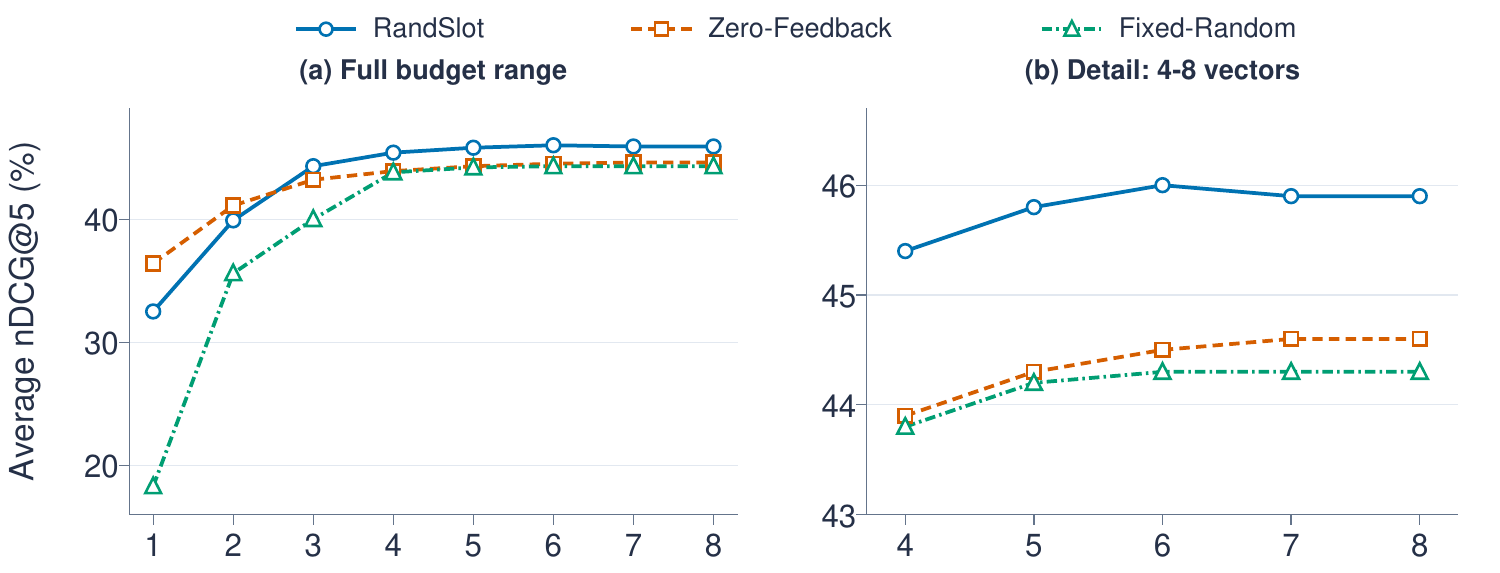}
\caption{Inference-budget scaling on ViDoRe V3 with Qwen2.5-VL-3B. Each method is trained with 4 vectors per side and evaluated with equal query and page budgets. The right panel enlarges the four-to-eight-vector range using a different vertical scale.} 
\label{fig:budget}
\end{figure}
We use Qwen2.5-VL-3B models trained with three appended auxiliary vectors per side, yielding four readouts in total: one from the original input and three from the appended positions. Fixed-Random reuses its third stored input at additional positions (Appendix~\ref{app:implementation}). Figure~\ref{fig:budget} reveals a crossover: Zero-Feedback leads at 1 and 2 vectors, whereas \method\ leads from 3 through 8. At 4 vectors, \method\ exceeds Zero-Feedback and Fixed-Random by 1.5 and 1.6 points, respectively. This sweep varies both sides together; we next isolate the contribution of each side.

Additional vectors yield diminishing returns. \method\ reaches 46.0 at 6 vectors and remains at 45.9 with 7 or 8, indicating that the useful budget saturates within the tested range. The model can therefore benefit from more readouts than it used during training, although further expansion does not consistently improve retrieval. Appendix~\ref{app:budget-cost} reports the complete numerical sweep.

\subsection{MaxSim winner assignments and complementarity}
\label{sec:readouts}
\begin{figure}[!t]
\centering
\includegraphics[width=\linewidth]{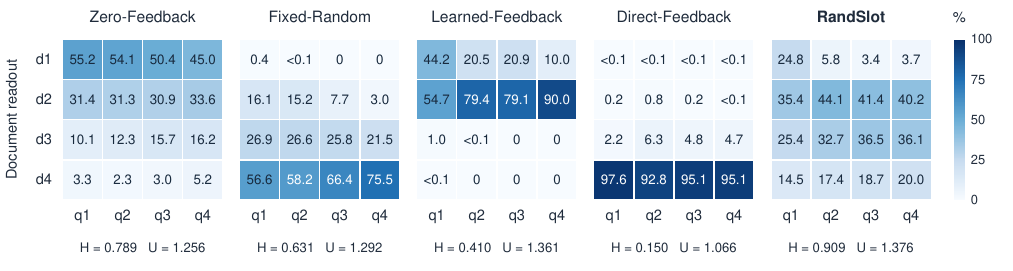}
\caption{MaxSim winner assignments across input strategies. For each positive query--page pair, each query readout $q_i$ selects the document readout $d_j$ with the largest cosine similarity. Columns sum to 100\% before rounding and all panels share a 0--100\% color scale. Results pool 74,016 positive pairs across the eight public ViDoRe V3 datasets. $H$ is normalized marginal winner entropy and $U$ is the mean number of distinct winners per pair.}
\label{fig:winner-assignments}
\end{figure}
For every positive pair, we record which document readout wins the MaxSim comparison for each query readout. Figure~\ref{fig:winner-assignments} shows two complementary summaries. The marginal entropy $H$ measures how evenly assignments are distributed over document positions across the pooled pairs, whereas $U$ measures how many distinct document positions are selected within an individual pair. \method\ has the highest $H$ (0.909) and the highest $U$ (1.376), compared with 0.789/1.256 for Zero-Feedback and 0.631/1.292 for Fixed-Random. Its assignments are therefore the least concentrated globally and use the broadest set of document readouts on average. The within-pair difference is modest---Learned-Feedback reaches $U=1.361$ despite a substantially lower $H$---so neither statistic alone establishes functional complementarity.

Directional MaxSim creates competition among document readouts: for each $q_i$, the matching path in Equation~\ref{eq:maxsim} passes through only the highest-scoring $d_j$. Resampling the appended inputs may perturb this competition across training examples and steps, reducing persistent dependence on a small set of document positions. RandSlot's more balanced winner distribution in Figure~\ref{fig:winner-assignments} provides empirical support for this interpretation, reflecting broader use of document readout positions at evaluation.

Broader matching assignments, together with stronger joint retrieval performance, help explain \method's effectiveness. The single-readout and removal experiments in Appendix~\ref{app:diagnostics} provide further support (Figure~\ref{fig:readout-contributions}). Although \method's best individual document readout is weaker than either control, its full set performs better, yielding a joint-use gain of 6.34 points, compared with 3.73 for Zero-Feedback and 3.17 for Fixed-Random. Together, these results suggest that resampling encourages the model to exploit complementary contributions from multiple document readouts. This collective use offers an explanation for how \method\ makes more effective use of a compact vector budget, even when individual readouts are not stronger on their own.

\subsection{Storage, matching, and encoding costs}
\label{sec:cost}
\begin{table}[!t]
\centering
\caption{Retrieval quality, theoretical representation costs, and measured encoding costs. Storage assumes 2,048 dimensions and two bytes per scalar for every row.}
\label{tab:encoding-cost}
\small
\setlength{\tabcolsep}{3pt}
\begin{tabular*}{\linewidth}{@{\extracolsep{\fill}}lrrrrrr@{}}
\toprule
 & & & \multicolumn{2}{c}{Theoretical costs} & \multicolumn{2}{c}{Measured encoding} \\
\cmidrule(lr){4-5}\cmidrule(lr){6-7}
Method & $K_q/K_d$ & V3 (\%) $\uparrow$ & KiB/page $\downarrow$ & Dot products $\downarrow$ & Query (ms) $\downarrow$ & Pages/s $\uparrow$ \\
\midrule
CausalEmbed$^\dagger$ & 16/32 & 42.6 & 128 & 512 & -- & -- \\
\method & 4/4 & 45.4 & 16 & 16 & 109.92 & 4.01 \\
\method & 6/6 & 46.0 & 24 & 36 & 162.08 & 3.29 \\
\bottomrule
\end{tabular*}
\par\smallskip
\begin{minipage}{\linewidth}\footnotesize $\dagger$: published score and budget from \citet{huo2026causalembed}. Storage covers raw page embeddings; dot products are counted per query--page pair. Query: median latency; Pages/s: page throughput. RandSlot uses one Qwen2.5-VL-3B checkpoint trained at 4/4. Scores follow the main results and budget sweep; timing uses a separate A800 run (BF16, batch one, KV cache, fixed input lengths; Appendix~\ref{app:diagnostics}). --: no measurement under our profiling protocol.\end{minipage}
\end{table}

Table~\ref{tab:encoding-cost} links retrieval quality to representation and encoding costs. At 2,048 dimensions and two bytes per scalar, four-vector \method\ uses 16\,KiB per page and 16 dot products per query--page pair, respectively one-eighth and one-thirty-second of CausalEmbed's published 16/32 budget, with a higher V3 score. These ratios measure representation costs, not end-to-end speedups (Appendix~\ref{app:budget-cost}).

We profile Qwen2.5-VL-3B on one A800-SXM4-80GB with BF16, batch size one, KV caching, 32 query tokens, and 1,536 merged visual tokens; Appendix~\ref{app:diagnostics} details the protocol and other feedback variants. Increasing the inference budget from 4 to 6 vectors raises V3 from 45.4 to 46.0, but separate profiling shows 47.4\% higher query latency and 17.9\% lower page throughput. Storage grows by 50\%, and dot products rise from 16 to 36. Four vectors provide a compact default; six trade higher costs for better quality.

\section{Discussion and Conclusion}
\label{sec:discussion}
\label{sec:conclusion}

We presented \method, a simple approach for compact visual document retrieval with resampled random unit soft tokens. Under a matched four-vector budget, \method\ outperforms alternative input strategies with two backbones and surpasses the published CausalEmbed results using substantially fewer vectors. Across seeds 42, 123, and 2024, the central 3B comparison preserves a 1.53-point average gain over Zero-Feedback. The gains also persist when random soft tokens are replaced with zeros at inference.

MaxSim assignments and readout interventions jointly reveal \method's advantage: matches span document readouts more broadly, and the full set delivers larger joint-use gains. These findings suggest that complementary use of multiple document readouts contributes to the observed gains. Document-side complementarity and query-side redundancy further reveal different representation needs, motivating separate query and document vector budgets.

Our evidence is limited to ViDoRe and two backbones. Due to compute and time constraints, additional seeds cover only the central 3B comparison, not every ablation, backbone, or benchmark. Broader repetitions and direct routing analyses remain future work. Overall, random soft tokens provide a simple and effective approach to compact visual document representation learning without introducing additional trainable token parameters.

\label{page:main-end}
\clearpage
\section*{Reproducibility statement}
\label{sec:reproducibility}
Section~\ref{sec:method} defines the readout construction and retrieval objective; Section~\ref{sec:experiments} describes the data, evaluation, and training protocol. Appendix~\ref{app:implementation} specifies the initial weights, input templates, readout positions, feedback construction and gradients, inference-budget extension rule, optimization, hardware, seeds, and checkpoint selection. Tables~\ref{tab:feedback} and~\ref{tab:train-infer} report the feedback and inference comparisons, with additional V3 details in Appendix~\ref{app:tasks}. Sections~\ref{sec:budget}--\ref{sec:cost} analyze the inference budget, readout contributions, and costs. Appendix~\ref{app:budget-cost} provides the full sweep and analytical costs; Appendix~\ref{app:diagnostics} specifies the separate diagnostic and timing protocols. The manuscript source includes the original score snapshot, the additional measurement records, and the table and figure generation scripts. Published results are explicitly attributed. We plan to release the training code and configuration files to support reproduction.

\section*{AI use statement}
\label{sec:ai-use}
We used generative AI assistance through OpenAI Codex to help refine research hypotheses and experimental comparisons, discuss interpretations of the observed results, and organize the conceptual framing and mathematical description of the method. We also used it for literature discovery and summarization, Chinese--English translation, manuscript drafting and revision, and preparation of figures, tables.

The quantitative results reported in this paper come from experimental records supplied by the authors or from explicitly cited publications. AI assistance was used to organize, compare, and describe these results. The authors directed the study and supplied and corrected the experimental records. The authors are responsible for reviewing the AI-assisted material and for the accuracy, originality, and integrity of the final manuscript.

\clearpage
\bibliography{references}
\bibliographystyle{plainnat}
\clearpage
\appendix
\section{Implementation details}
\label{app:implementation}
Table~\ref{tab:configuration} summarizes the optimization and execution settings shared by the two backbones. We detail the initialization, input construction, and feedback implementations below.

\begin{table}[!t]
\centering
\caption{Training and execution settings for Qwen2.5-VL-3B and Qwen3-VL-4B. Both backbones use the same training hyperparameters.}
\label{tab:configuration}
\small
\begin{tabular}{@{}p{0.29\linewidth}p{0.67\linewidth}@{}}
\toprule
Item & Configuration \\
\midrule
Output budget & Four vectors per query and page \\
Output dimensions & 2,048 (3B); 2,560 (4B); no output projection \\
Training length & One epoch; 923 steps \\
Checkpoint & Final checkpoint after step 923 \\
Hardware & Eight NVIDIA A100 80GB GPUs per training run \\
Batch & 128 global; 16 per device; gradient accumulation 1 \\
Forward microbatch & 1 \\
LoRA & Rank 16, scaling 32, dropout 0.05; Q/K/V/O projections \\
Optimizer & PagedAdamW8bit; $\beta=(0.9,0.999)$, $\epsilon=10^{-8}$, weight decay 0.01 \\
Learning rate & $2\times10^{-4}$ \\
Schedule & Cosine branch; $T_{\max}=2\times923=1{,}846$ steps \\
Gradient clipping & Maximum norm 1.0 \\
Retrieval objective & Query-to-document InfoNCE; MaxSim; temperature 0.07 \\
Contrastive candidates & Document representations gathered across devices \\
Visual-token limit & 1,536 during training and evaluation \\
Computation & BF16, FlashAttention 2, gradient checkpointing \\
Training execution & Sequential appending; full-prefix recomputation \\
Evaluation execution & KV caching enabled \\
Random seed & 42 for training and inference \\
\bottomrule
\end{tabular}
\end{table}

\paragraph{Backbones and initialization.}
We initialize from \nolinkurl{Qwen/Qwen2.5-VL-3B-Instruct} and \nolinkurl{Qwen/Qwen3-VL-4B-Instruct}. Within each backbone, all feedback variants start from the same pretrained weights, and query and page encoders share parameters. Retrieval vectors are L2-normalized last-layer hidden states without an output projection, giving dimensions of 2,048 and 2,560, respectively.

\paragraph{Training data and query inputs.}
We use the \texttt{train} split of \texttt{vidore/colpali\_train\_set} without additional filtering or subsampling. The original \texttt{query} field and its paired page image form each positive pair. Queries receive no additional instruction prefix, role marker, or explicitly appended end token. Instructions already present in the query field are retained; the query augmentation string is empty.

\paragraph{Page inputs.}
Both backbones use the following concatenated strings as the page prompt, where \texttt{\textbackslash n} denotes a newline:
\begingroup
\footnotesize
\begin{verbatim}
(
    "<|im_start|>user\n"
    "<|vision_start|><|image_pad|><|vision_end|>"
    "Describe the image.<|im_end|><|endoftext|>"
)
\end{verbatim}
\endgroup
The processor expands the image placeholder into visual-token positions. No assistant role or generation prompt is added.

\paragraph{Padding and readout positions.}
We use left padding, and the first readout comes from the last text token of a query or the terminal \texttt{<|endoftext|>} token of a page. At the default budget, the other three readouts come from appended embedding positions. Auxiliary vectors are appended directly to \texttt{inputs\_embeds}, without discrete token IDs. All outputs are L2-normalized before MaxSim scoring.

\paragraph{Readout execution and random sampling.}
The implementation appends auxiliary embeddings sequentially and reads the final output of each extended prefix. Training recomputes the full prefix, retaining previously appended random values within the same encoding and drawing only for the next position. Each new encoding samples an independent sequence; draws are independent across examples, modalities, and appended positions. Standard Gaussian sampling and L2 normalization use FP32 before conversion to the computation dtype. Evaluation enables KV caching to reuse the prefix and process each new position.

\paragraph{Fixed and learned feedback tables.}
Learned-Feedback retains MetaEmbed's learnable-token input mechanism but not its nested-budget objective, instead using our unified retrieval objective, training recipe, and four-vector budget. Fixed-Random and Learned-Feedback use separate position-indexed query and page tables, each containing three vectors at the default budget. Fixed-Random initializes random unit vectors once and preserves them through training, checkpoint loading, and inference. Learned-Feedback uses randomly initialized \texttt{nn.Embedding} parameters, which are optimized directly. LoRA is applied only to the backbone encoder, not to these input embeddings. Embedding updates use PagedAdamW8bit with a learning rate of $2\times10^{-4}$ and the same optimizer hyperparameters as the backbone's LoRA parameters (Table~\ref{tab:configuration}). Each learned vector is L2-normalized before appending; its parameters are updated during training and reused at inference. Learned-Feedback adds $2(K-1)D_{\rm in}$ trainable scalars for $K$ outputs per side. Table entries are shared across examples, while contextualized outputs depend on the input. Every feedback variant trains the encoder adaptation parameters.

\paragraph{Direct-Feedback gradients.}
Direct-Feedback retains CausalEmbed's autoregressive reinsertion mechanism but omits its method-specific auxiliary losses, instead using our unified retrieval objective, training recipe, and four-vector budget. It reinserts the current normalized retrieval vector as the next input, $\mathbf e_t=\mathbf r_t$. We neither detach this vector nor stop gradients along the reinsertion path, allowing subsequent readouts to backpropagate through the feedback inputs into earlier computations.

\paragraph{Fixed-Random beyond the training budget.}
Let $\mathbf f_1^{(m)},\mathbf f_2^{(m)},\mathbf f_3^{(m)}$ be the stored vectors for modality $m$. For an inference budget of $K$ outputs, Fixed-Random uses
\begin{equation}
\mathbf e_t^{(m)}=\mathbf f_{\min(t,3)}^{(m)},\qquad t=1,\ldots,K-1.
\label{eq:fixed-extension}
\end{equation}
The first three appended positions use the stored vectors in order; all additional positions repeat the third vector. No new vectors are sampled, and the table is not cycled. The first output still reads the original input, and encoder weights remain fixed during the sweep.

\paragraph{Checkpoint selection and reporting.}
Each configuration is trained for one epoch, comprising 923 steps, and evaluated using the final checkpoint after step 923. Unless stated otherwise, the reported results use seed 42 and correspond to a single run. For the central Qwen2.5-VL-3B comparison between Zero-Feedback and RandSlot, we evaluate across seeds 42, 123, and 2024; Appendix~B.2 reports the mean and sample standard deviation. Because of the substantial computational cost, we restrict repeated runs to this central comparison and do not repeat every feedback ablation, backbone, or benchmark setting.


\clearpage
\section{Additional ViDoRe V3 results}
\label{app:tasks}
\subsection{Feedback strategies with Qwen2.5-VL-3B}
Table~\ref{tab:tasks-3b} reports the eight ViDoRe V3 tasks for the five feedback strategies. The largest gains over Zero-Feedback are on Industrial ($+2.6$) and HR ($+2.4$), while Pharmaceuticals and Fin-F decline by 0.3 and 0.6 points, respectively.
\begin{table}[H]
\centering
\caption{ViDoRe V3 feedback comparisons with Qwen2.5-VL-3B (nDCG@5, \%). R, Z, F, L, and DF denote RandSlot, Zero-Feedback, Fixed-Random, Learned-Feedback (Separate), and Direct-Feedback. Differences are in percentage points.}
\label{tab:tasks-3b}
\begingroup
\small
\setlength{\tabcolsep}{4pt}
\begin{tabular}{@{}lrrrrrr@{}}
\toprule
Task & R & Z & F & L & DF & $\Delta$(R--Z) \\
\midrule
HR & 44.1 & 41.7 & 42.5 & 42.2 & 40.8 & +2.4 \\
Fin-E & 44.0 & 42.0 & 41.9 & 37.3 & 41.4 & +2.0 \\
Ind. & 35.3 & 32.7 & 33.7 & 31.9 & 30.0 & +2.6 \\
Phar. & 49.1 & 49.4 & 47.4 & 50.3 & 48.4 & -0.3 \\
C.S. & 67.3 & 65.1 & 64.5 & 64.2 & 62.1 & +2.2 \\
Ener. & 49.8 & 48.5 & 48.0 & 48.3 & 45.7 & +1.3 \\
Phys. & 43.6 & 41.8 & 43.6 & 42.0 & 40.0 & +1.8 \\
Fin-F & 29.8 & 30.4 & 28.4 & 27.0 & 28.2 & -0.6 \\
\bottomrule
\end{tabular}
\endgroup
\end{table}

\subsection{Robustness across random seeds}
\label{app:seed-robustness}
To assess whether the Table~\ref{tab:feedback} gain is sensitive to training randomness, we evaluate the Qwen2.5-VL-3B Zero-Feedback and RandSlot comparison across seeds 42, 123, and 2024. Table~\ref{tab:seed-robustness} reports the mean and sample standard deviation over the three complete training and evaluation runs. \method\ improves the macro-average nDCG@5 from $43.50\!\pm\!0.43$ to $45.03\!\pm\!0.40$, a gain of $1.53$ points, and improves seven of the eight subsets. The similar run-level variability supports that the gain is not specific to the original seed. Owing to computational constraints, we conduct this multi-seed check only for the central 3B Zero-Feedback comparison; the remaining ablations, the 4B backbone, and the V1/V2 settings retain their single-run results.
\begin{table}[H]
\centering
\caption{Seed robustness of the Qwen2.5-VL-3B comparison on ViDoRe V3 (nDCG@5 $\uparrow$, \%). Values are mean $\pm$ sample standard deviation over seeds 42, 123, and 2024. $\Delta$ is the difference between the two means in percentage points.}
\label{tab:seed-robustness}
\begingroup
\small
\setlength{\tabcolsep}{5pt}
\begin{tabular*}{\linewidth}{@{\extracolsep{\fill}}lccc@{}}
\toprule
Dataset & Zero-Feedback & \method & $\Delta$ \\
\midrule
Computer Science & $63.34 \pm 1.50$ & $\mathbf{66.54 \pm 0.84}$ & $+3.20$ \\
Industrial       & $32.48 \pm 0.22$ & $\mathbf{34.24 \pm 0.83}$ & $+1.76$ \\
Pharmaceuticals  & $49.59 \pm 0.42$ & $\mathbf{49.63 \pm 0.66}$ & $+0.04$ \\
Physics          & $41.60 \pm 0.36$ & $\mathbf{43.83 \pm 0.75}$ & $+2.22$ \\
Energy           & $48.72 \pm 0.79$ & $\mathbf{50.14 \pm 0.65}$ & $+1.42$ \\
Finance-EN       & $41.65 \pm 0.66$ & $\mathbf{42.45 \pm 1.66}$ & $+0.80$ \\
Finance-FR       & $\mathbf{30.06 \pm 1.01}$ & $29.11 \pm 0.82$ & $-0.96$ \\
Human Resources  & $40.58 \pm 1.24$ & $\mathbf{44.33 \pm 0.22}$ & $+3.75$ \\
\midrule
\textbf{Macro Average} & $43.50 \pm 0.43$ & $\mathbf{45.03 \pm 0.40}$ & $\mathbf{+1.53}$ \\
\bottomrule
\end{tabular*}
\endgroup
\end{table}

\newpage
\subsection{Inference feedback with Qwen2.5-VL-3B}
Table~\ref{tab:tasks-3b-zero} details the V3 inference replacement experiment for \method-3B. Random and zero inference use the same randomly trained encoder and both average 45.4. Under zero-input inference, random training improves six tasks over zero training and lowers two, showing the task-level variation behind the retained average gain.
\begin{table}[H]
\centering
\caption{ViDoRe V3 inference feedback replacement with Qwen2.5-VL-3B (nDCG@5, \%). RS-R and RS-Z share the randomly trained encoder; Z is trained with zero feedback. The final column compares zero inference after random and zero training, in percentage points.}
\label{tab:tasks-3b-zero}
\begingroup
\small
\setlength{\tabcolsep}{6pt}
\begin{tabular}{@{}lrrrr@{}}
\toprule
Task & RS-R & Z & RS-Z & $\Delta$(RS-Z--Z) \\
\midrule
HR & 44.1 & 41.7 & 44.1 & +2.4 \\
Fin-E & 44.0 & 42.0 & 44.2 & +2.2 \\
Ind. & 35.3 & 32.7 & 35.0 & +2.3 \\
Phar. & 49.1 & 49.4 & 49.0 & -0.4 \\
C.S. & 67.3 & 65.1 & 67.5 & +2.4 \\
Ener. & 49.8 & 48.5 & 49.8 & +1.3 \\
Phys. & 43.6 & 41.8 & 43.6 & +1.8 \\
Fin-F & 29.8 & 30.4 & 29.7 & -0.7 \\
\bottomrule
\end{tabular}
\endgroup
\end{table}

\subsection{Feedback and inference with Qwen3-VL-4B}
Table~\ref{tab:tasks-4b} reports the V3 feedback and inference results for \method-4B. \method\ averages 46.1, compared with 45.1 for both Zero-Feedback and Fixed-Random and 44.4 for Learned-Feedback. Its largest gains over Zero-Feedback occur on C.S. ($+4.6$) and Industrial ($+3.2$), while Pharmaceuticals, Fin-E, and Fin-F decline by 1.7, 0.9, and 0.8 points.

Direct-Feedback averages 45.4, exceeding Zero-Feedback, Fixed-Random, and Learned-Feedback. \method\ improves five of the eight tasks over Direct-Feedback, with the largest gains on C.S. ($+3.3$) and Physics ($+1.7$). Fin-E, Pharmaceuticals, and Fin-F decline by 0.2, 0.7, and 1.6 points, respectively, yielding an average gain of 0.7 points.

Zero-input inference retains a V3 average of 45.8, remaining above the zero-trained model's 45.1. Against fixed and learned feedback, the largest gains from random inference are on Fin-E ($+4.7$ and $+4.9$, respectively). Differences are in percentage points, computed at the displayed precision.
\begin{table}[H]
\centering
\caption{ViDoRe V3 feedback and inference results with Qwen3-VL-4B (nDCG@5, \%). R and RS-Z use random and zero inference after random training; Z uses zero feedback throughout. F and L use fixed-random and learned modality-specific tables; DF reinserts the latest readout. Each $\Delta$ column gives R minus the indicated method, in percentage points. Averages retain the reported precision.}
\label{tab:tasks-4b}
\begingroup
\footnotesize
\setlength{\tabcolsep}{2pt}
\begin{tabular*}{\linewidth}{@{\extracolsep{\fill}}lrrrrrrrrrr@{}}
\toprule
Task & R & Z & F & L & DF & RS-Z & $\Delta$Z & $\Delta$F & $\Delta$L & $\Delta$DF \\
\midrule
HR & 44.6 & 43.9 & 43.1 & 42.6 & 44.0 & 44.8 & +0.7 & +1.5 & +2.0 & +0.6 \\
Fin-E & 44.9 & 45.8 & 40.2 & 40.0 & 45.1 & 44.4 & -0.9 & +4.7 & +4.9 & -0.2 \\
Ind. & 35.2 & 32.0 & 34.7 & 33.2 & 33.7 & 34.7 & +3.2 & +0.5 & +2.0 & +1.5 \\
Phar. & 51.2 & 52.9 & 51.5 & 51.4 & 51.9 & 50.9 & -1.7 & -0.3 & -0.2 & -0.7 \\
C.S. & 70.8 & 66.2 & 70.1 & 69.2 & 67.5 & 70.5 & +4.6 & +0.7 & +1.6 & +3.3 \\
Ener. & 49.9 & 48.8 & 49.9 & 47.9 & 49.5 & 49.2 & +1.1 & +0.0 & +2.0 & +0.4 \\
Phys. & 44.8 & 43.6 & 43.5 & 43.7 & 43.1 & 44.6 & +1.2 & +1.3 & +1.1 & +1.7 \\
Fin-F & 27.1 & 27.9 & 27.8 & 27.4 & 28.7 & 26.9 & -0.8 & -0.7 & -0.3 & -1.6 \\
\midrule
Average & 46.1 & 45.1 & 45.1 & 44.4 & 45.4 & 45.8 & +1.0 & +1.0 & +1.7 & +0.7 \\
\bottomrule
\end{tabular*}
\endgroup
\end{table}

\clearpage
\section{Evaluation summaries and cost accounting}
\label{app:budget-cost}
\begin{table}[!t]
\centering
\caption{Inference-budget results on ViDoRe V3 (nDCG@5, \%). Each method uses a fixed Qwen2.5-VL-3B model trained with four vectors per side, with equal query and page budgets at inference.}
\label{tab:budget}
\begingroup
\footnotesize
\setlength{\tabcolsep}{3.3pt}
\begin{tabular}{@{}lrrrrrrrr@{}}
\toprule
Vectors per side $i$ & 1 & 2 & 3 & 4 & 5 & 6 & 7 & 8 \\
\midrule
RandSlot & 32.5 & 39.9 & 44.3 & 45.4 & 45.8 & 46.0 & 45.9 & 45.9 \\
Zero-Feedback & 36.4 & 41.1 & 43.2 & 43.9 & 44.3 & 44.5 & 44.6 & 44.6 \\
Fixed-Random & 18.3 & 35.6 & 40.0 & 43.8 & 44.2 & 44.3 & 44.3 & 44.3 \\
\bottomrule
\end{tabular}
\endgroup
\end{table}

\begin{table}[!t]
\centering
\caption{ViDoRe V3 quality (nDCG@5, \%) and theoretical representation costs. Storage assumes 2,048 dimensions and two bytes per scalar for every row.}
\label{tab:efficiency}
\begingroup
\footnotesize
\setlength{\tabcolsep}{3pt}
\begin{tabular*}{\linewidth}{@{\extracolsep{\fill}}lrrrrrr@{}}
\toprule
Method & $K_q$ & $K_d$ & V3 & KiB/page & GiB/100K & Dot products \\
\midrule
CausalEmbed$^\dagger$ & 16 & 32 & 42.6 & 128 & 12.21 & 512 \\
RandSlot (3B) & 4 & 4 & 45.4 & 16 & 1.53 & 16 \\
RandSlot (3B) & 6 & 6 & 46.0 & 24 & 2.29 & 36 \\
\bottomrule
\end{tabular*}
\endgroup
\par\smallskip
\begin{minipage}{\linewidth}\footnotesize $\dagger$: published score and vector budget from \citet{huo2026causalembed}. 100K denotes 100,000 pages; dot products are counted per query--page pair. Both RandSlot rows use one checkpoint trained at 4/4. Costs exclude index overhead and encoding time.\end{minipage}
\end{table}

\paragraph{Score reporting.}
We report nDCG@5 on a 0--100 scale and absolute differences in percentage points. Main results use one decimal place and readout diagnostics use two, preserving the original precision after scaling. Source records remain on the 0--1 scale; version averages are not recomputed from rounded task scores.

\paragraph{Complete inference-budget results.}
Table~\ref{tab:budget} provides the values plotted in Figure~\ref{fig:budget}. Each strategy uses a Qwen2.5-VL-3B model trained with four vectors per side. The model is then held fixed while query and page budgets vary together from one to eight; no model is retrained for individual inference budgets.

\paragraph{Analytical representation costs.}
For $N$ pages, $K_d$ page vectors, dimension $D$, and $b$ bytes per scalar, the raw embedding payload is
\begin{equation}
S_{\rm page}=K_dDb,\qquad S_{\rm corpus}=NK_dDb.
\end{equation}
Exact MaxSim evaluates $K_qK_d$ dot products of dimension $D$ per query--page pair. The dot-product stage requires $K_qK_dD$ multiply--accumulate operations, followed by maximum and summation reductions. Tables~\ref{tab:encoding-cost} and~\ref{tab:efficiency} count the dot products and apply $D=2{,}048$ and $b=2$ to compare representation budgets. These calculations exclude encoding, index structures, and other system overhead; the dot-product ratios are not measured retrieval speedups.

Under these assumptions, four vectors require 16\,KiB per page and approximately 1.53\,GiB for 100,000 pages; six require 24\,KiB and 2.29\,GiB. CausalEmbed's 32 document vectors correspond to 128\,KiB and 12.21\,GiB. KiB and GiB denote $1{,}024$ and $1{,}024^3$ bytes. All feedback strategies have the same payload and theoretical matching cost at a fixed budget, dimension, and precision. Only the final retrieval vectors are stored, not the auxiliary soft-token inputs.

\clearpage
\section{Readout diagnostics and encoding measurements}
\label{app:diagnostics}
\begin{figure}[H]
\centering
\includegraphics[width=\linewidth]{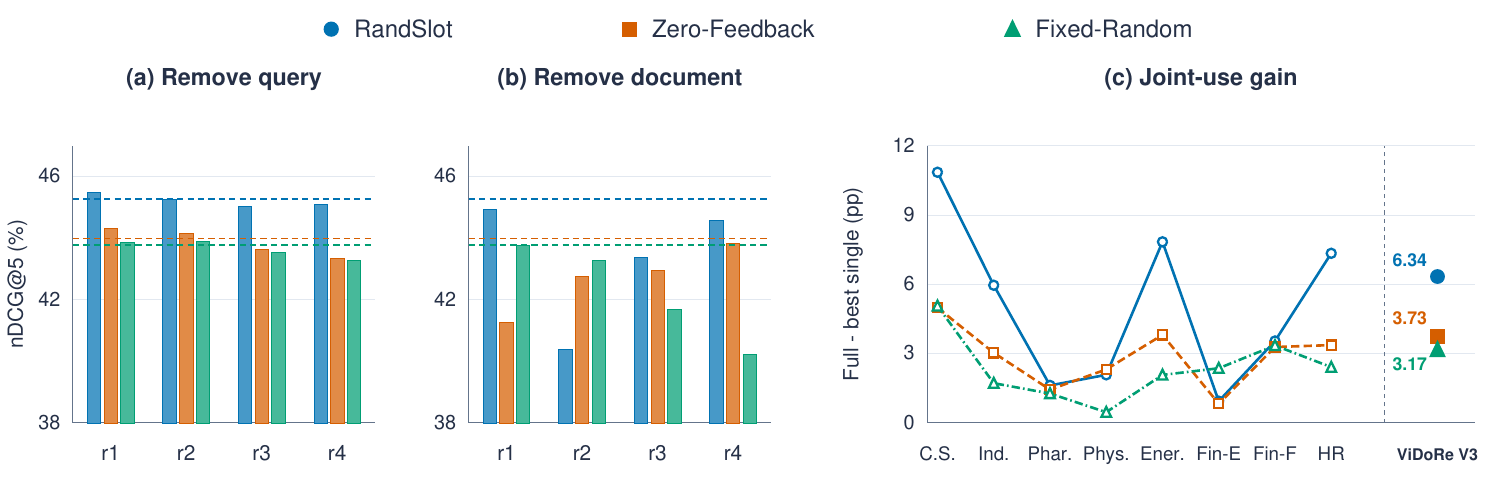}
\caption{Readout contributions on ViDoRe V3 (nDCG@5, \%). (a--b) Scores after removing a query/document readout; dashed lines denote full sets (axes: 38--47). (c) Full minus best-single-document scores. Subsets select their best position separately; \textbf{ViDoRe V3} selects one fixed position by macro score, not the mean of preceding gains. The opposite side retains four readouts. Gains are in percentage points (pp). Lines guide the eye.}
\label{fig:readout-contributions}
\end{figure}

\paragraph{Diagnostic evaluation.}
We use the Qwen2.5-VL-3B checkpoints for Zero-Feedback, Fixed-Random, and \method, each trained with four outputs per side. The eight public V3 tasks contain 14,514 queries and 19,252 pages; all queries have relevance judgments. Encoding uses eight devices, batch size 32, and evaluation seed 42; scoring uses FP32. Scores are averaged equally across tasks. This separate diagnostic run supplies only the within-run readout comparisons in Figure~\ref{fig:readout-contributions} and Table~\ref{tab:readout-details}; it does not replace the main results or the original budget sweep.

\begin{table}[H]
\centering
\caption{Readout diagnostics on ViDoRe V3. Scores are percentages; decreases are percentage points. The opposite side retains all four vectors. Single: score using only $r_i$ on the indicated side; removal: full score minus score without $r_i$. Negative values indicate improvement after removal.}
\label{tab:readout-details}
\footnotesize
\setlength{\tabcolsep}{2pt}
\begin{tabular*}{\linewidth}{@{\extracolsep{\fill}}llrrrrrrrr@{}}
\toprule
 & & \multicolumn{4}{c}{Single readout: nDCG@5 (\%)} & \multicolumn{4}{c}{Removal: decrease (pp)} \\
\cmidrule(lr){3-6}\cmidrule(lr){7-10}
Method & Side & $r_1$ & $r_2$ & $r_3$ & $r_4$ & $r_1$ & $r_2$ & $r_3$ & $r_4$ \\
\midrule
RandSlot & Query & 41.42 & 44.92 & 45.42 & 45.27 & -0.22 & -0.01 & 0.21 & 0.17 \\
RandSlot & Document & 34.96 & 38.92 & 37.73 & 33.97 & 0.33 & 4.88 & 1.89 & 0.69 \\
\midrule
Zero-Feedback & Query & 39.97 & 42.81 & 44.29 & 44.94 & -0.34 & -0.17 & 0.36 & 0.63 \\
Zero-Feedback & Document & 40.26 & 38.32 & 36.43 & 33.32 & 2.71 & 1.22 & 1.04 & 0.14 \\
\midrule
Fixed-Random & Query & 40.85 & 42.26 & 43.84 & 43.90 & -0.10 & -0.14 & 0.23 & 0.49 \\
Fixed-Random & Document & 20.00 & 36.07 & 38.74 & 40.60 & 0.00 & 0.48 & 2.07 & 3.55 \\
\bottomrule
\end{tabular*}
\end{table}

\paragraph{Winner-assignment aggregation.}
Figure~\ref{fig:winner-assignments} pools 74,016 positive query--page pairs across the eight public V3 datasets and is therefore pair-weighted rather than an equal-weight dataset macro-average. Retrieval readouts are L2-normalized before scoring. For each query readout, the four document-winner frequencies sum to one before rounding. The supplied diagnostics report no ties under a tolerance of $10^{-8}$; ties would otherwise be resolved by the lowest index. The normalized marginal entropy is $H=-\sum_j p_j\log p_j/\log 4$, where $p_j$ is the pooled winner frequency of document position $j$. The statistic $U$ averages the number of distinct winning document positions among the four query readouts for each positive pair.

\paragraph{Scoring interventions.}
After full encoding, we cache the vectors and retain or remove a selected readout on one side during MaxSim scoring. The opposite side always retains four vectors. Removal decreases are computed against the full-set score from the same run. A single query readout therefore still matches four document vectors. No feedback input or forward step is removed, so these interventions do not reduce encoding work. The observed position preferences are diagnostics, not an independently validated pruning policy.

Figure~\ref{fig:readout-contributions}(a--b) shows an asymmetric pattern. Removing any \method\ document readout lowers the macro score, most strongly for $r_2$ (4.88 points) and $r_3$ (1.89 points). On the query side, removing $r_1$ improves the score by 0.22 points, and using only $r_3$ exceeds the full query set by 0.16 points. Both controls also exhibit query-side redundancy.

\paragraph{Joint-use gains.}
Let $F_s$ be the full-document score on subset $s$ and $S_{s,i}$ the score using only document readout $i$, with all four query readouts retained. The subset points in Figure~\ref{fig:readout-contributions}(c) report $100(F_s-\max_i S_{s,i})$. The final ViDoRe V3 column reports $100(\overline{F}-\max_i\overline{S_i})$, where bars denote equal-weight means over the eight subsets. The selected fixed positions are $r_2$, $r_1$, and $r_4$ for \method, Zero-Feedback, and Fixed-Random, respectively. Averaging the subset-specific gains instead yields 5.03, 2.89, and 2.34 points. Position selection is retrospective on the evaluation results and serves only as a diagnostic reference.

In the readout diagnostic analysis, \method's best fixed document readout performs worse than those of Zero-Feedback and Fixed-Random, whereas its full set outperforms both controls. Combining all document readouts also yields a larger gain over the best fixed individual readout for \method\ than for either control. With subset-specific best positions, \method\ gains on all eight subsets and exceeds each control's gain on seven, except Physics against Zero-Feedback and English Finance against Fixed-Random. These interventions support document-side functional complementarity without proving semantic specialization or a causal role for geometric diversity.

\paragraph{Timing protocol.}
We profile one NVIDIA A800-SXM4-80GB with BF16, batch size one, KV caching, PyTorch 2.6.0+cu124, and CUDA 12.4. Each configuration uses three warm-up samples and 32 measured samples with CUDA synchronization. For profiling only, queries are fixed to 32 active tokens by repeating short inputs and retaining the final tokens; pages are resized to $48\times32$ merged visual tokens. Timings include host-to-device transfer and model forward passes, excluding preprocessing and retrieval. They measure fixed-shape encoding, not end-to-end retrieval under natural input lengths.

\paragraph{Additional encoding measurements.}
Under the same four-vector profiling protocol, Zero-Feedback takes 108.48\,ms per query at 3.99 pages/s, Fixed-Random takes 105.74\,ms at 4.05 pages/s, and zero-input \method\ takes 107.60\,ms at 4.02 pages/s. These aggregate measurements come from a single profiling run; small differences among same-budget variants do not establish a consistent speed advantage.

\end{document}